# Diet variation in climbing perch populations inhabiting eight different types of ecosystems

V. V. BINOY[1] AND P. S. PRASANTH[2]

[1]National Institute of Advanced Studies, Indian Institute of Science Campus, Bangalore 560 012, India

[2] Palakkath House, Vellangallur, Thrissur District, Kerala 680 662, India

e-mail: vvbinoy@gmail.com

Tel.: +91 782 949 6778; fax: +91 802 218 5028;

Running title: Cross population variation in the diet of climbing perch

ABSTRACT

Present study revealed that populations of climbing perch (*Anabas testudineus*) inhabiting river, backwater, shallow water channel, ponds with and without vegetation cover, marsh, sewage canal and aquaculture tank varied significantly in the number of food items consumed. Chironomous larvae, organic debris and filamentous algae were the common ingredients of the menu of this species across focal ecosystems, whereas sewage canal population was found surviving solely on insect larvae and organic debris.

Keywords: Climbing perch, Population variation, Cross-ecosystem variation, Gut content analysis, Feeding behaviour

Populations of a piscine species inhabiting ecosystems markedly divergent in the biotic and abiotic properties can provide valuable insights into the mechanisms of adaptation and evolution, since each ecosystem imparts different kinds of selection pressures on the individuals. In many contexts the divergence in the ecological parameters is reflected in the quality and quantity of food items available in a particular habitat and individuals from contrasting ecosystems may have to consume food materials demanding dissimilar handling strategies and physiological mechanisms for digestion. Food being one of the major determinant of the fitness of a species in an ecosystem knowledge of the cross population variation in the foraging strategies and food items consumed is essential for the conservation and restocking of a species that could survive in the contrasting habitat conditions as well as to prevent the migration of invasive piscine species to new areas.

The climbing perch (*Anabas testudineus*), a common freshwater species of India and other south east Asian countries, being equipped with accessory respiratory organ can survive in a wide variety of lenthic and lotic habitats including the once that are hostile (sewage or brackish water) to other species (Prasanth, 2006). Hence, this fish is an excellent model system to understand the ecosystem-induced divergence in various traits including foraging behaviours. Additionally, in Indian subcontinent natural populations of this species with good market demand are facing decline due to overexploitation and habitat loss (Kohinoor et al., 2012) while in Australia this species is raising serious ecological concern by becoming an alien invasive species (Hitchcock, 2008). Even after being a species of ecological and economical importance, unfortunately very few studies has focused the cross ecosystem variation in feeding behaviour of this species till date. The present study analyzed variation in the diet of climbing perch population surviving in eight different habitats significantly diverse in ecological properties.

Climbing perch was collected from following eight different ecosystems (Fig.1) located in different parts of Thrissur District, Kerala state (Prasanth, 2006) with the help of expert fishermen during the pre-monsoon period. The ecosystems focused were, River (Karuvannur River (S1); 10.39°N; 76.22°E), Backwater (tributary of Karupadanna River (S2); 10.21°N; 76.21°E), shallow water channels (depth 1.5 ± 0.5 m) in banana plantation (Thirumukulam village (S3); 10.53°N; 76.30°E) and a pond with dense covering vegetation comprising *Hydrilla verticillata* and *Pistia stratiotes* (Vellangallur village (S4); 10.30°N; 76.22°E). Fish was also collected from a small pond without any vegetation cover (Athani (S5); 10.28°N; 76.22°E), canal draining murky and foul smelling sewage (Irinjalakuda (S6); 10.33°N; 76.21°E), and a marsh (Irinjalakuda (S7); 10.35°N; 76.22°E). In order to check whether climbing perch kept in aquaculture tanks consume any plankton in addition to the artificial pellets provided, individuals kept in outdoor cement tank (depth 1.75 m and diameter 1.5 m) of an aquarium was also collected. This tank (Thumboor (S8); 10.29°N; 76.25°E) with extensive algal growth had a substratum composed of decaying mahogany (*Swietenia mahogani*) leaves and was housing 72 climbing perches. These fishes were fed only with commercially available artificial food pellets.

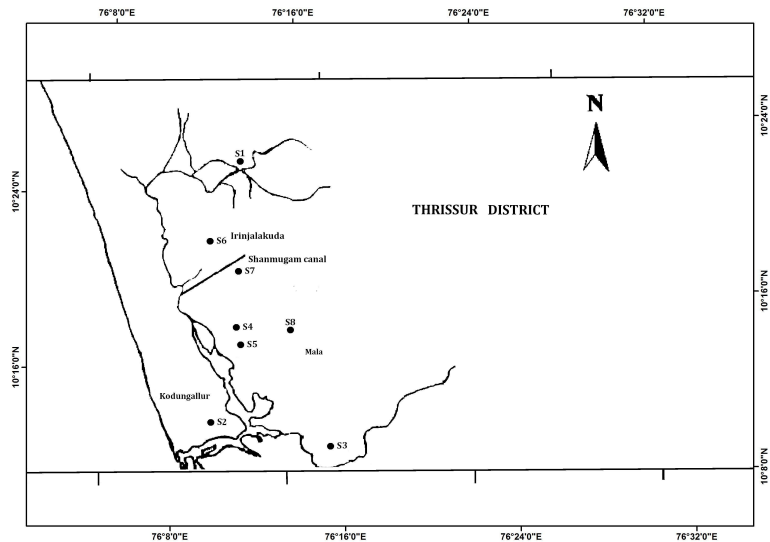

Fig. 1. Different ecosystem from which climbing perch was collected for studying diet variation.

The fish collected from each ecosystem were sacrificed under deep anesthesia and each individual were injected with 5% formalin solution and preserved immediately in ice to minimize further digestion of food items in the alimentary canal. Gut content of ten randomly selected individuals from each ecosystem were conducted in the laboratory after dissection.

Although climbing perch had been described as a predator, carnivore (Pandey *et al.,* 1992) or an insectivore (Ahyaudin, 1992) by many authors, presence of algae, decaying plant matter, parts of aquatic invertebrates, small fishes etc. in the gut content of the individuals collected from different ecosystems (Table 1) support the suggestion that this species is omnivorous nature (Nargis and Hossain, 1987). However in contrary to the study of Nargis and Hossain (1987), which implies that that diet of climbing perch is more or less consistent irrespective of spatial and seasonal variation, our result revealed a significant variation in the number of food items consumed by climbing perch in different ecosystems (ANOVA $F_{7,80}$ = 37.39; p<0.001; Fig. 2). Further analysis of the data using Tukeys test (Table 2) revealed that number of food items consumed by climbing perch from sewage canal and culture tank was significantly less in comparison to other populations. The climbing perch was solely dependent on *Chironomous* larvae and organic debris in the sewage canal and the amount of food materials present in the gut of the fish collected from this habitat was least in contrast to other population. Whereas the fish from marsh was found to be consuming maximum diversity of food items in comparison to the individuals from all other ecosystems except pond with vegetation cover and river.

| **Food items** | **Ecosystems** | | | | | | | |
|---|---|---|---|---|---|---|---|---|
| | R | B | S | M | Pv | Pnv | Sw | A |
| Small fish | * | | | * | | | | |
| *Chironomous sp.* larva | * | | * | | | * | * | * |
| Insect larva | | | | * | * | | | |
| Small shrimps | | * | | | | | | |
| Aquatic arthropods | | | * | | * | | | |
| Decaying soft plant materials | * | * | * | * | * | * | * | * |
| Pieces of wood | | | | | * | | | |
| *Oscillatoria sp* | * | * | | | * | | | |
| *Spyrogyra sp* | | | | * | | | | |
| Organic debris | * | * | * | * | * | * | * | * |
| Artificial pellets | | | | | | | | * |

Table 1. Gut content of climbing perch collected from different kinds of ecosystems. River (R), Backwater (B), Shallow channels (S), Marsh (M), Pond with vegetation cover (Pv), Small pond without vegetation cover (Pnv) Sewage canal (Sw), Artificial tanks (A).

Another interesting observation was the presence of Cyanophycean filamentous alga *Oscillatoria sp.* in the diet of individuals inhabiting river, backwater and shallow water channel habitats (Prasanth, 2006). Moreover this alga was the major food item for climbing perch living in the pond with vegetation cover, while individuals from marsh were found to be dependent up on another filamentous alga *Spirogyra sp.* Interestingly, even though the water was bloomed with non filamentous species, no alga was found in the stomach of climbing perch reared in the artificial tanks. According to Vivekanandan *et al.,* (1977) addition of filamentous alga *Sprorogyra maxima* in the meat based diet provided to climbing perch in laboratory condition could significantly reduce metabolic load and hence the frequency of air gulping behaviour. Moreover, Panserata *et al.,* (2009) has shown that changes in the food items consumed can distinctly modify the expression of genes controlling vital physiological functions in fish. Hence the need and effect of consuming different species of filamentous algae on the metabolism and behaviour of this species and its impact on the fitness of populations surviving in different ecosystems offers future avenues for research. Additionally, although some authors (Potongkam, 1972; Nargis and Hossain, 1987) suggest mollusks as an ingredient of climbing perch diet, our study failed to find signs of the consumption of gastropods by any individual collected during the study.

|    | B     | S     | M     | Pv    | Pnv   | Sw       | C         |
|----|-------|-------|-------|-------|-------|----------|-----------|
| R  | -1.60 | -1.28 | 1.92  | 0.64  | -1.60 | -9.28*** | -9. 28*** |
| B  |       | 0.32  | 3.52* | 2.24  | 0     | -7.68*** | -7.68***  |
| S  |       |       | 3.20* | 1.92  | -0.32 | -8.00*** | -8.00***  |
| M  |       |       |       | -1.28 | -3.52*| 11.20*** | -11.20*** |
| Pv |       |       |       |       | -2.24 | 9.92***  | -9.92***  |
| Pnv|       |       |       |       |       | -7.69*** | -7.68***  |
| Sw |       |       |       |       |       |          | 0         |

Table 2. *Post hoc* analysis of the number of food items consumed by climbing perch in contrasting ecosystems. The statistics used is Tukey's test.

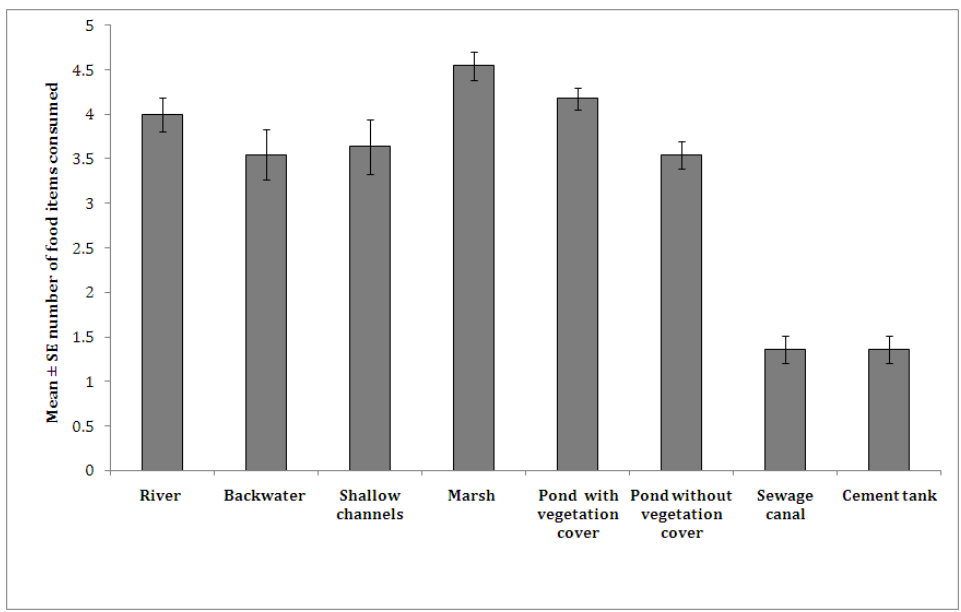

Fig. 2. Cross-ecosystem variation in the number of food items present in the gut of climbing perch.

To conclude, present study revealed that climbing perch exhibit cross-ecosystem variation in the diet and individuals of this species are surviving only on a few kind of food items in the sewage canal. Additionally presence of algae in the gut content of fish collected from many of the ecosystems studied points to the need for including this item in the diet of climbing perch grown in artificial setups. Here, it is important to note that climbing perch is famous for its tendency to migrate from one habitat to another during monsoon period and very little literature is available describing how individuals living in one ecosystem will be coping up, physiologically and behaviorally, if it reaches a contrasting ecosystem. Hence, the result of future studies comparing genetic diversity, physiology, morphology, behaviour and coping strategies of climbing perch populations surviving in diverse ecosystems will elucidate the secret behind successful survival of climbing perch in a wide range of habitats.

## Acknowledgement

VVB gratefully acknowledges Science Engineering Research Board, Department of Science and Technology, Government of India for funding this research under the Start-up Research Grant (Young Scientist). Authors are grateful to C V Nandini, KUFOS, Kochi, for her assistance in preparing map of study area.

## REFERENCES

Ahyaudin, B. A.  1992.  Rice farming development in Malaysia past, present and future, In: Rice fish research and development in Asia. *Proceedings of ICLARM Conference, Manila*, Philippines, p. 257.

Hitchcock, G.  2008. Climbing perch (*Anabas testudineus*)(Perciformes: Anabantidae) on Saibai Island, northwest Torres Strait: First Australian record of this exotic pest fish. *Mem. Queensl. Muse.*, 52: 207–211.

Kohinoor, A. H. M., Islam, M. S., Jahan, D. A., Khan, M. M. and Hussain, M. G. 2012. Growth and production performances of crossbred climbing perch koi, *Anabas testudineus* in Bangladesh. *Int. J. Agril. Res. Innov. Tech.* 2: 19–25

Nargis, A. and Hossain, M. A. 1987. Food and feeding habit of Koi fish (*Anabas testudineus* Bloch. Anabantidae: Perciformes). *Bangladesh J. Agri.,* 12(2): 121-127.

Pandey, A., Srivastava, P. K., Adirkari, S. and Singh, D. K. 1992. pH Profile of gut as an index of food and feeding habits of fishes. *J. Freshwat. Biol.,* 4(2): 75-79.

Panserata, S., Hortopand, G. A., Plagnes-Juana, E., Kolditza, C., Lansarda, M., Skiba-Cassya, S. Esquerrée, D., Geurdena, I., Médalea, F., Kaushika, S. and Corrazea, G. 2009. Differential gene expression after total replacement of dietary fish meal and fish oil by plant products in rainbow trout (*Oncorhynchus mykiss*) liver*. Aquaculture,* 294: 123-131.

Potongkam, K. 1972. Experiment on feeding climbing perch, *Anabas testudineus* (Bloch) with ground trash fish and pellets. *Annual Report, Department of Fisheries, Bangkok, Thailand.*

Prasanth, P. S. 2006. Influence of variations in habitat on the feeding behavior and reproductive strategies of climbing perch (*Anabas testudineus)* a freshwater fish. M.Sc. Dissertation, University of Calicut, India.

Vivekanandan, E., Pandian, T. J. and Visalam, C. N. 1977. Effects of algal and animal food combinations on surfacing activity and food utilization in the climbing perch *Anabas scandens. Pol. Arch. Hydrobiol.,* 24 (4): 555-562.